\documentstyle[aps,epsf,manuscript]{revtex}

\begin{document}

\newcommand{\half}{\frac12}
\newcommand{\vare}{\varepsilon}
\newcommand{\eps}{\epsilon}
\newcommand{\pr}{^{\prime}}
\newcommand{\ppr}{^{\prime\prime}}
\newcommand{\pp}{{p^{\prime}}}
\newcommand{\hp}{\hat{\bfp}}
\newcommand{\hpp}{\hat{\bfpp}}
\newcommand{\hq}{\hat{\bfq}}
\newcommand{\rqq}{{\rm q}}
\newcommand{\rx}{{\rm x}}
\newcommand{\rp}{{\rm p}}
\newcommand{\rpp}{{{\rm p}^{\prime}}}
\newcommand{\rk}{{\rm k}}
\newcommand{\bfp}{{\bf p}}
\newcommand{\bfpp}{{\bf p}^{\prime}}
\newcommand{\bfq}{{\bf q}}
\newcommand{\bfx}{{\bf x}}
\newcommand{\bfk}{{\bf k}}
\newcommand{\bfz}{{\bf z}}
\newcommand{\bnabla}{{\mbox{\boldmath$\nabla$}}}
\newcommand{\bphi}{{\mbox{\boldmath$\phi$}}}
\newcommand{\balpha}{{\mbox{\boldmath$\alpha$}}}
\newcommand{\bsigma}{{\mbox{\boldmath$\sigma$}}}
\newcommand{\bomega}{{\mbox{\boldmath$\omega$}}}
\newcommand{\bvare}{{\mbox{\boldmath$\varepsilon$}}}
\newcommand{\intzo}{\int_0^1}
\newcommand{\intinf}{\int^{\infty}_{-\infty}}
\newcommand{\ka}{\kappa_a}
\newcommand{\kb}{\kappa_b}
\newcommand{\lbr}{\langle}
\newcommand{\rbr}{\rangle}
\newcommand{\ThreeJ}[6]{
        \left(
        \begin{array}{ccc}
        #1  & #2  & #3 \\
        #4  & #5  & #6 \\
        \end{array}
        \right)
        }
\newcommand{\SixJ}[6]{
        \left\{
        \begin{array}{ccc}
        #1  & #2  & #3 \\
        #4  & #5  & #6 \\
        \end{array}
        \right\}
        }
\newcommand{\NineJ}[9]{
        \left\{
        \begin{array}{ccc}
        #1  & #2  & #3 \\
        #4  & #5  & #6 \\
        #7  & #8  & #9 \\
        \end{array}
        \right\}
        }
\newcommand{\Dmatrix}[4]{
        \left(
        \begin{array}{cc}
        #1  & #2   \\
        #3  & #4   \\
        \end{array}
        \right)
        }
\newcommand{\cross}[1]{#1\!\!\!/}
\newcommand{\beq}{\begin{equation}}
\newcommand{\eeq}{\end{equation}}
\newcommand{\beqn}{\begin{eqnarray}}
\newcommand{\eeqn}{\end{eqnarray}}

%%%%%%%%%%%%%%%%%%%%%%%%%%%%%%%%%%%%%%%%%%%%%%%%%%%%%%%%%%%%%%%%%%%%%%%%%
%
%%%%%%%%%%%%%%%%%%%%%%%%%%%%%%%%%%%%%%%%%%%%%%%%%%%%%%%%%%%%%%%%%%%%%%%%%
%
\title{
Self-energy correction to the bound-electron $g$ factor in H-like ions}
\author{
V. A. Yerokhin$^{1,2}$, P. Indelicato$^1$, and V. M. Shabaev$^{2,3}$ }

\address{
$^1$ Laboratoire Kastler-Brossel, \'Ecole Normale Sup\'erieure et Universit\'e Pierre et
Marie Curie, Case 74, 4 place Jussieu, F-75252, Cedex 05, France\\
$^2$ Department of Physics, St. Petersburg State University, Oulianovskaya 1, St.
Petersburg 198504, Russia \\
$^3$ Gesellschaft f\"ur Schwerionenforschung, 64291 Darmstadt, Germany
}
\date{\today}
\maketitle
  
\begin{abstract}
The one-loop self-energy correction to the $1s$ electron $g$ factor is evaluated to all
orders in $Z\alpha$ with an accuracy, which is essentially better than that of previous
calculations of this correction. As a result, the uncertainty of the theoretical
prediction for the bound-electron $g$ factor in H-like carbon is reduced by a factor of
3. This improves the total accuracy of the recent electron-mass determination [Beier
{\it et al.} Phys. Rev. Lett. {\bf 88}, 011603 (2002)]. The new value of the electron
mass is found to be $m_e = 0.000\, 548\, 579\, 909\, 3\, (3)$ u.

\noindent PACS number(s): 31.30.Jv, 12.20.Ds
\end{abstract}
\vspace*{0.5cm}
%%%%%%%%%%%%%%%%%%%%%%%%%%%%%%%%%%%%%%%%%%%%%%%%%%%%%%%%%%%%%%%%%%%%%%%%%
%
%%%%%%%%%%%%%%%%%%%%%%%%%%%%%%%%%%%%%%%%%%%%%%%%%%%%%%%%%%%%%%%%%%%%%%%%%
%
Spectacular progress in high-precision measurements of the bound-electron $g$ factor
for the hydrogen-like carbon \cite{Hermanspahn00,Haeffner00} and the related
theoretical investigations recently provided a new independent determination of the
electron mass \cite{Beier02}. It yields $$ m_e(g) = 0.000\, 548\, 579\, 909\, 2\, (4)\
\mbox{\rm u}\,.$$ This result agrees with the 1998 CODATA value \cite{Mohr00} $$
m_e(\mbox{\rm CODATA}) = 0.000\, 548\, 579\, 911\, 0\, (12)\ \mbox{\rm u}\,$$ within
1.5 standard deviations but is three times more precise. The uncertainty of the
electron-mass value of \cite{Beier02} originates equally from the theoretical result
for the bound-electron $g$ factor and from the experimental value for the ratio of the
electronic Larmor precession frequency and the cyclotron frequency of the ion in the
trap. Therefore, any advance in theoretical or experimental investigations will
improve the accuracy of the electron-mass value. However, for significant progress one
needs to reduce both the theoretical and experimental uncertainties. From the
experimental side, an increase of the accuracy by an order of magnitude is anticipated in
the near future, as well as an extension of the measurements to higher-$Z$ systems
\cite{Werth01}. Investigations of the bound-electron $g$ factor in high-$Z$ systems
are of particular importance since they can provide a new determination of the fine
structure constant \cite{Karshenboim00,Werth01}, nuclear magnetic moments
\cite{Werth01}, and nuclear charge radii. They would also create a good possibility
for testing the magnetic sector of QED in a strong Coulomb field.

From the theoretical point of view, the leading error of the bound-electron g-factor
value for H-like carbon comes from the one-loop self-energy correction. Reducing this
uncertainty is the aim of the present investigation. The second major error is due to the
two-loop binding QED correction that is known at present only to the lowest order in
$Z\alpha$ \cite{Karshenboim00,cza01,Grotch70}. 
To reduce that uncertainty is a serious problem. However,
recent progress in calculations of two-loop QED corrections to the Lamb shift, both
within the $Z\alpha$ expansion \cite{Pachucki01} and to all orders in $Z\alpha$
\cite{Yerokhin01}, allows us to hope that its solution might be possible in the near future.
An important feature of studying the bound-electron $g$ factor is a relative
weakness of nuclear effects. Unlike the hyperfine splitting, where a large effect of
distribution of the magnetic moment over the nucleus complicates the identification of
one-loop QED effects, for the bound-electron $g$ factor, the uncertainty due to
nuclear effects is of the order of two-loop binding QED corrections even in the
high-$Z$ region. In addition, as shown in \cite{Shabaev02a}, the finite-nuclear-size
effect can be largely cancelled in a specific difference of the bound-electron $g$
factors for H- and Li-like ions with the same nucleus. Therefore, this difference can be (in principle)
calculated up to a very high accuracy. This fact makes the bound-electron $g$ factor
very promising for testing two-loop QED effects by comparing theory and
experiment.

The one-loop self-energy correction to the $1s$ $g$ factor was first evaluated by
Blundell {\it et al.} \cite{Blundell97} and by Persson {\it et al.} \cite{Persson97}.
The latter work was extended by Beier and co-workers \cite{Beier00}, whose result was
used in the electron-mass determination \cite{Beier02}.

Formal expressions for the one-loop self-energy correction to the bound-state g
factor are well-known (see, e.g., \cite{Beier00a}). The whole correction is
conveniently divided into three parts, which are referred to as the irreducible
($\Delta g_{\rm ir}$), the reducible ($\Delta g_{\rm red}$) and the vertex ($\Delta
g_{\rm ver}$) contribution. The irreducible part is given by
\beq \label{eq1}
\Delta g_{\rm ir} = \frac2b  \lbr a | \Sigma_R(\vare_a) |\delta a\rbr\,,
\eeq
where $\Sigma_R$ denotes the renormalized self-energy operator, and $a$ indicates the
initial state. We use the relativistic units ($\hbar = c = 1$) and the Heaviside
charge unit [$\alpha = e^2/(4\pi)$, $e < 0$] throughout this Letter. The perturbed
wave function $|\delta a\rbr$ is
\beq \label{eq2}
|\delta a\rbr\ = \sum_{\vare_n \ne \vare_a} \frac{|n\rbr \lbr n| \delta V|
                                                       a\rbr}{\vare_a-\vare_n}\,,
\eeq
where $\delta V(\bfx) = -e \balpha \cdot {\bf A}_{\rm cl}(\bfx)$, ${\bf A}_{\rm cl}$
denotes the classical homogeneous magnetic field, ${\bf A}_{\rm cl}(\bfx) = [ {\cal
H}\times \bfx] /2$, $b$ is defined by $\Delta E = b\, \Delta g$, $b = \mu_0 m_a {\cal
H}$, $\mu_0 = |e|/(2m)$ is the Bohr magneton, and $m_a$ is the angular-momentum
projection of the initial state. The reducible contribution is represented as
\beq    \label{eq3}
\Delta g_{\rm red} = \frac1b \lbr a| \delta V|a\rbr
          \lbr a |\left. \frac{\partial}{ \partial \vare} \Sigma_R(\vare)
                          \right|_{\vare = \vare_a}| a\rbr\,,
\eeq
and the vertex part is given by
\beq    \label{eq4}
\Delta g_{\rm ver} = \frac1b \sum_{n_1n_2} \frac{i}{2\pi} \intinf d\omega\,
   \frac{\lbr n_1| \delta V|n_2 \rbr \lbr a n_2 |I(\omega)| n_1 a\rbr}
     {[\vare_a-\omega-\vare_{n_1}(1-i0)][\vare_a-\omega-\vare_{n_2}(1-i0)]}\,,
\eeq
where $I(\omega) = e^2 \alpha^{\mu} \alpha^{\nu} D_{\mu \nu}(\omega)$ is the operator
of the electron-electron interaction, $D_{\mu \nu}(\omega)$ stands for the photon
propagator, and $\alpha^{\mu} =  (1,\balpha)$ are the Dirac matrices. In order to
avoid large numerical cancellations, it is convenient to calculate the vertex and the
reducible part together. We indicate the sum of these two contributions with the
subscript "{\rm vr}", $\Delta g_{\rm vr} = \Delta g_{\rm ver}+\Delta g_{\rm red}$.

Now we turn to the numerical evaluation of these contributions. We perform our
calculations in the Feynman gauge and both for the point and the extended nucleus. In
the latter case, the hollow-shell nuclear model was utilized. Since calculations for
the point nucleus are easier from the technical point of view and because of smallness
of the finite-nuclear-size effect, we later discuss mainly the point-nucleus evaluation.
Convergence of the extended-nucleus value to the point-nucleus result
for small values of $Z$ served as one of the checkups for our numerical procedure. The
calculation of the irreducible part is quite straightforward. For the point nucleus,
the perturbed wave function $|\delta a \rbr$ can be found analytically by employing
the generalized virial relations for the Dirac equation \cite{Shabaev91}. The
corresponding explicit expressions can be found in \cite{Shabaev01}. The numerical
evaluation of the non-diagonal matrix element of the self-energy operator was carried
out similarly to that for the self-energy correction to the hyperfine structure
\cite{Yerokhin01a}, within the Green-function technique. The partial-wave expansion
converges well in that case, and taking into account 30-50 partial waves is sufficient
for getting the required accuracy (with the rest of the series estimated by polynomial
fitting). As an additional cross-check for the evaluation of the irreducible part, we
also utilized a modified renormalization procedure, where the energy of the
one-potential term is shifted from its physical value (for more details we refer the
reader to \cite{Yerokhin01a}).

The numerical evaluation of the vertex and reducible parts is more problematic. The
standard way to treat corrections of this kind is to separate terms in which bound
electron propagators are replaced with free propagators. We refer to this part as the
{\it 0-potential} contribution $\Delta g_{\rm vr}^{(0)}$. This term contains
ultraviolet divergences that can be covariantly separated and cancelled in momentum
space. The remainder $\Delta g_{\rm vr}^{(1+)}$ is ultraviolet finite and can be
calculated directly in coordinate space, as in \cite{Blundell97}. However, it turns
out that due to a strong cancellation between the reducible and the vertex part, the
contribution of high partial waves is relatively large for low $Z$, and the
corresponding expansion is slowly converging. For gaining better control over the
partial-wave summation, in \cite{Persson97,Beier00} it was proposed to separate from
$\Delta g_{\rm vr}^{(1+)}$ a part containing (besides an interaction with the magnetic
field) one Coulomb interaction with the nucleus in electron propagators, the so-called
{\it 1-potential} contribution $\Delta g_{\rm vr}^{(1)}$. The authors demonstrated
that the partial-wave expansion of the remainder (the {\it many-potential}
contribution $\Delta g_{\rm vr}^{(2+)}$) is converging much better than that for
$\Delta g_{\rm vr}^{(1+)}$. For the evaluation of the 1-potential term, a separate
numerical scheme was developed in \cite{Persson97,Beier00}, based on an analytical
treatment of radial integrals. This allowed the authors to extend the partial-wave
summation up to $l_{\rm max} = 120$. However, the unevaluated tail of the expansion
still yields a significant contribution in that case. In order to get the accuracy,
ascribed to the 1-potential term in \cite{Beier00} for carbon, one should estimate the
tail of the series with an uncertainty of about 1\%. This is a potentially dangerous
point of this numerical evaluation.

The central point of the present calculation is a different treatment of the 1-potential
term. We evaluate it directly in momentum space without utilizing the partial-wave
expansion and, in this way, eliminate the uncertainty due to the estimation of the tail of
the series. The next difference from the calculations \cite{Persson97,Beier00}
consists in the treatment of the magnetic interaction in momentum space. The Fourier
transform of the classical magnetic potential involves the gradient of a $\delta$
function,
\beq \label{eq5}
{\bf A}_{\rm cl}(\bfq) = -\frac{i}{2}(2\pi)^3 [{\cal H}\times \bnabla_{\bfq}
   \delta^3(\bfq)] \,.
\eeq
In \cite{Persson97,Beier00}, the $\delta$ function was replaced by a continuous
Gaussian function with a small but finite regulator. In our evaluation of the 0- and
1-potential terms, we employ directly (\ref{eq5}) and evaluate the corresponding
corrections after integration by parts. (For the 0-potential term, the same approach was
utilized earlier in \cite{Blundell97}.) In case of the 0-potential term, this
treatment requires additional analytical work, but finally, instead of
a five-dimensional numerical integration (as in \cite{Persson97,Beier00}), we end up
with a single integral that can be evaluated up to an arbitrary precision. The analytical
part of the evaluation of the 1-potential term is quite tedious, but the overall
$\delta$ function simplifies the calculation greatly. Finally, the 1-potential term is
represented by a four-dimensional integral, whose numerical evaluation is relatively
easy.

The calculation of the many-potential term was carried out in a 
manner similar  to that in \cite{Yerokhin01a}. The many-potential part was represented by a
point-by-point difference of the unrenormalized, the 0-potential, and the 1-potential
term. In addition, we  also subtract the infrared-divergent contribution of the
reference state from the vertex and reducible parts. 
This contribution was then evaluated separately, carrying out the
$\omega$ integration analytically and explicitly cancelling divergences in the sum of
the reducible and the vertex part. Care should be taken in the evaluation of the
many-potential correction, since a large numerical cancellation occurs in the
point-by-point difference. In order to avoid the appearance of pole terms that lead to
additional numerical cancellations, we employ the following contour of the $\omega$
integration: $(\vare_0-i\infty, \vare_0-i0] + [\vare_0-i0,-i0] +[i0,\vare_0+i0]
+[\vare_0+i0,\vare_0+i\infty)$, rather than simply the integration over the imaginary
axis. The parameter $\vare_0$ in the definition of the contour can be varied. In
actual calculations its value was taken to be about $Z\alpha\, \vare_a$ for low $Z$. The
summation over partial waves was carried out up to $|\kappa_{\rm max}| = 20$-$35$, and
the tail of the series was estimated by polynomial fitting.

The results of our numerical evaluation are presented in Table I. In order to isolate
the one-loop binding self-energy correction, we subtract from the total self-energy
correction the free-electron value $\alpha/\pi$ \cite{Schwinger48}. The resulting
binding correction is compared with the data from \cite{Beier00}. For all cases except
for $Z=20$, the results agree with each other within the given error bars. A more
detailed comparison is presented in Table II for two most important cases, carbon
and oxygen. A certain deviation can be observed for the 1-potential and
many-potential contributions, that is largely cancelled in the sum. We do not have any
explanation of this fact at present. As an additional checkup of our calculation, we
fitted our data for the binding correction and compared the result for the leading
$(Z\alpha)^2$ term with the analytical value \cite{Grotch70} $a_{20} = 1/6 =
0.1666\ldots$. Our fitting yields $a_{20} = 0.1667(2)$. Finally, we separate the
higher-order contribution $F_{\rm h.o.}(Z\alpha)$ that incorporates terms of order
$(Z\alpha)^4$ and higher,
\beq \label{eq6}
\Delta g_{\rm SE} = \frac{\alpha}{\pi} \left[ 1 + \frac16 (Z\alpha)^2 + (Z\alpha)^4
   F_{\rm h.o.}(Z\alpha) \right]\,.
\eeq
The results for the higher-order contribution are represented in Fig. 1, together with
those from \cite{Beier00}. A least-squares fit of our data to the form
\beq
F_{\rm h.o.}(Z\alpha) = a_{41} \ln (Z\alpha)+ a_{40}+ (Z\alpha)[ \cdots ]
\eeq
yields $a_{41} = -7.0(8)$ and $a_{40} = -10(2)$.

In Table III we present individual contributions to the $1s$ electron $g$ factor for two most
important cases, H-like carbon and oxygen. The Dirac point-nucleus value and the
free-electron part of the one-loop QED correction are evaluated utilizing the
recommended value for the fine-structure constant from \cite{Mohr00}. The 
finite-nuclear-size correction is calculated numerically and is in good agreement with
the previous evaluations \cite{Beier00,Glazov02}. 
The so-called "electric-loop" part of the one-loop vacuum-polarization correction 
is also re-evaluated in this work. The corresponding results agree with 
the earlier numerical 
\cite{Persson97,Beier00} and analytical \cite{Karshenboim01} calculations.
The remaining part of the vacuum-polarization
("magnetic-loop") correction is shown to be negligible for the case under
consideration \cite{Karshenboim01}. The
$\alpha^2$ QED correction includes the existing $Z \alpha$ expansion terms for the QED
correction of second order in $\alpha$ \cite{Karshenboim00,cza01} and the known
free-QED terms of higher orders in $\alpha$ (see, e.g., \cite{Beier00a}). Its relative
uncertainty was estimated as the ratio of the part of the one-loop QED correction
that is beyond the $(Z \alpha)^2$ approximation, to the part that is within the
$(Z\alpha)^2$ approximation, multiplied by a factor of 1.5. The recoil correction incorporates
the total recoil contribution of first order in $m/M$, calculated to all orders in
$Z\alpha$ in \cite{Shabaev02}, and the known corrections of orders $(m/M)^2$ and
$\alpha(m/M)$ \cite{fau00}.

In summary, our evaluation of the one-loop self-energy correction for the $1s$
electron $g$ factor in H-like ions improves the accuracy of the theoretical prediction
for carbon by a factor of 3 and for oxygen by a factor of 2. This reduces the total
uncertainty of the electron-mass determination of \cite{Beier02}. The new value for
the electron mass is found to be
\beq
m_e = 0.000\, 548\, 579\, 909\, 29\, (29)(8)\,,
\eeq
where the first uncertainty originates from the experimental value for the ratio of
the electronic Larmor precession frequency and the cyclotron frequency of the ion in
the trap, and the second error comes from the theoretical value for the bound-electron
$g$ factor.

We would like to thank Th. Beier for valuable discussions and, in particular, for
communicating the details of the extrapolation procedure of the partial-wave expansion
used in his calculation. Valuable conversations with H.-J. Kluge and W. Quint are
gratefully acknowledged.  
This study was supported in part by the Russian Foundation
for Basic Research (project no. 01-02-17248), 
by the program "Russian Universities" (project no. UR.01.01.072), and by GSI.
The stay of V.Y. in Paris is supported by the Minist\`ere de 
l'Education Nationale et de la Recherche.
Laboratoire Kastler Brossel is Unit\'e Mixte de Recherche de l'\'Ecole 
Normale Sup\'erieure, du CNRS et de l'Universit\'e P. et M. Curie.

%%%%%%%%%%%%%%%%%%%%%%%%%%%%%%%%%%%%%%%%%%%%%%%%%
%%% Table I
%%%%%%%%%%%%%%%%%%%%%%%%%%%%%%%%%%%%%%%%%%%%%%%%%
\begin{table}
\squeezetable
\caption{
Various contributions to the one-loop self-energy correction to the $1s$ electron
$g$ factor for H-like ions. All values are absolute contributions to the $g$ factor
($1/\alpha = 137.035\,989\,5$) and presented in units of $10^{-6}$ (ppm). The
point and the extended nuclear model are indicated with labels "pnt." and "ext.",
respectively. The binding correction is obtained by subtraction of the free-electron
value $\alpha/\pi$ from the total self-energy correction $\Delta g_{\rm SE}$.
Only the total numerical error of the present evaluation is
indicated.
}
\vspace*{0.5cm}
\begin{tabular}{r|r@{.}lr@{.}lr@{.}lr@{.}l|r@{.}l|r@{.}lr@{}lr@{.}l}
$Z$ & \multicolumn{2}{c}{$\Delta g_{\rm ir}$} &
                    \multicolumn{2}{c}{$\Delta g_{\rm vr}^{(0)}$} &
                                    \multicolumn{2}{c}{$\Delta g_{\rm vr}^{(1)}$}
                                                   & \multicolumn{2}{c}{$\Delta g_{\rm vr}^{(2+)}$}
                                                                &\multicolumn{2}{c}{ $\Delta g_{\rm SE}$}
                                                                                   &\multicolumn{2}{c}{ Binding}
                                                                                                    & \multicolumn{2}{c}{Binding}
                                                                                                                       &\multicolumn{2}{c}{ Ref. \cite{Beier00}} \\
     & \multicolumn{2}{c}{(pnt.)}
                   & \multicolumn{2}{c}{(pnt.)}
                                    & \multicolumn{2}{c}{(pnt.)}
                                                   &  \multicolumn{2}{c}{(pnt.)}
                                                                & \multicolumn{2}{c}{(pnt.)}
                                                                                   & \multicolumn{2}{c}{(pnt.)}
                                                                                                    &  \multicolumn{2}{c}{(ext.)}
                                                                                                                        & \multicolumn{2}{c}{(ext.)} \\
             \hline
 1   &     1&52923 &    2320&77563  &     0&50250  &    0&03305 &   2322&84041(10) &  0&02078(10)   &     &             &  0&0208(9)   \\
 2   &     5&20641 &    2316&00970  &     1&55757  &    0&13053 &   2322&90421(10) &  0&08458(10)   &     &             &  0&0844(9)   \\
 3   &    10&52313 &    2309&28506  &     2&91759  &    0&28869 &   2323&01447(10) &  0&19484(10)   &     &             &  0&1944(9)   \\
 4   &    17&21614 &    2300&99753  &     4&45945  &    0&50260 &   2323&17572(10) &  0&35609(10)   &     &             &  0&3555(9)   \\
 5   &    25&10745 &    2291&41521  &     6&10392  &    0&76661 &   2323&39319(10) &  0&57356(10)   &     &             &  0&5732(9)   \\
 6   &    34&06468 &    2280&73799  &     7&79535  &    1&07460 &   2323&67262(10) &  0&85299(10)   &     &             &  0&8528(9)   \\
 8   &    54&78171 &    2256&69788  &    11&16571  &    1&79701 &   2324&44231(10) &  1&62268(10)   &    1&.62267(10)    &  1&6225(10)  \\
10   &    78&74380 &    2229&82629  &    14&34905  &    2&61754 &   2325&53668(12) &  2&71705(12)   &    2&.71702(12)    &  2&7159(10)  \\
12   &   105&51170 &    2200&79829  &    17&21623  &    3&48376 &   2327&00998(15) &  4&19035(15)   &    4&.19030(15)    &  4&1907(12)  \\
15   &   150&23525 &    2154&28732  &    20&77184  &    4&75746 &   2330&05187(20) &  7&23224(20)   &    7&.23212(20)    &  7&231(1)  \\
18   &   199&76448 &    2105&29703  &    23&34254  &    5&85654 &   2334&26059(25) & 11&44096(25)   &   11&.44067(25)    & 11&442(2)   \\
20   &   235&17646 &    2071&71454  &    24&49971  &    6&41815 &   2337&80886(30) & 14&98923(30)   &   14&.98870(30)    & 15&04(1)     \\
\end{tabular}
\end{table}

%%%%%%%%%%%%%%%%%%%%%%%%%%%%%%%%%%%%%%%%%%%%%%%%%
%%% Table II
%%%%%%%%%%%%%%%%%%%%%%%%%%%%%%%%%%%%%%%%%%%%%%%%%

\begin{table}
\caption{
Various contributions to the one-loop self-energy correction to the $1s$ electron
$g$ factor for H-like carbon and oxygen in the present evaluation and  by
Beier {\it et al}.
Units are ppm. Our numerical values correspond to the point
nuclear model. Due to the smallness of the finite-nuclear-size effect, a different
treatment of the nucleus in the two calculations does not influence the
term-by-term comparison.
}
\begin{tabular}{l|r@{.}lr@{.}lr@{.}lr@{.}l|r@{.}l}
                      & \multicolumn{2}{c}{$\Delta g_{\rm ir}$} 
                                    & \multicolumn{2}{c}{$\Delta g_{\rm vr}^{(0)}$} 
                                                   & \multicolumn{2}{c}{$\Delta g_{\rm vr}^{(1)}$}
                                                                & \multicolumn{2}{c}{$\Delta g_{\rm vr}^{(2+)}$} 
                                                                             & \multicolumn{2}{c}{$\Delta g_{\rm SE}$} \\
             \hline
$Z$=6, this work      & 34&06468(4) & 2280&73799   & 7&79535(2) &  1&07460(9) & 2323&67262(10) \\
$Z$=6, \cite{Beier00} & 34&0647(4)  & 2280&7380(3) & 7&7945(1)  &  1&0752(1)  & 2323&6724(9) \\
             \hline
$Z$=8, this work      & 54&78171(4) & 2256&69788   & 11&16571(2)&  1&79701(9) & 2324&44231(10) \\
$Z$=8, \cite{Beier00} & 54&7815(4)  & 2256&6979(3) & 11&1646(2) &  1&7981(1)  & 2324&4421(10)\\
\end{tabular}
\end{table}

%%%%%%%%%%%%%%%%%%%%%%%%%%%%%%%%%%%%%%%%%%%%%%%%%
%%% Table III
%%%%%%%%%%%%%%%%%%%%%%%%%%%%%%%%%%%%%%%%%%%%%%%%%

\begin{table}
\caption{The $1s$ electron $g$ factor in H-like carbon and oxygen.}
\begin{tabular}{lr@{.}lr@{.}l}
              &  \multicolumn{2}{c}{$^{12}{\rm C}^{5+}$}
                                &   \multicolumn{2}{c}{$^{16}{\rm O}^{7+}$}\\
                    \hline
Dirac value (point)        &   1&998 721 354 4   &  1&997 726 003 1  \\
Fin. nucl. size            &   0&000 000 000 4   &  0&000 000 001 5  \\
QED, order $(\alpha/\pi)$  &   0&002 323 663 9(1)&  0&002 324 415 6(1)\\
QED, order $(\alpha/\pi)^2$&  -0&000 003 516 2(3)& -0&000 003 517 1(6)\\
Recoil                     &   0&000 000 087 6   &  0&000 000 117 0  \\
Total                      &   2&001 041 590 1(3)&  2&000 047 020 1(6)\\
\end{tabular}
\end{table}

%%%%%%%%%%%%%%%%%%%%%%%%%%%%%%%%%%%%%%%%%%%%%%%%%%%%%%%%%%%%%%%%%%%%%%%%
%%%%%
%%%%%
%%%%%%%%%%%%%%%%%%%%%%%%%%%%%%%%%%%%%%%%%%%%%%%%%%%%%%%%%%%%%%%%%%%%%%%
%\newpage
\begin{figure}
\centerline{ \mbox{
\epsfxsize=0.9\textwidth \epsffile{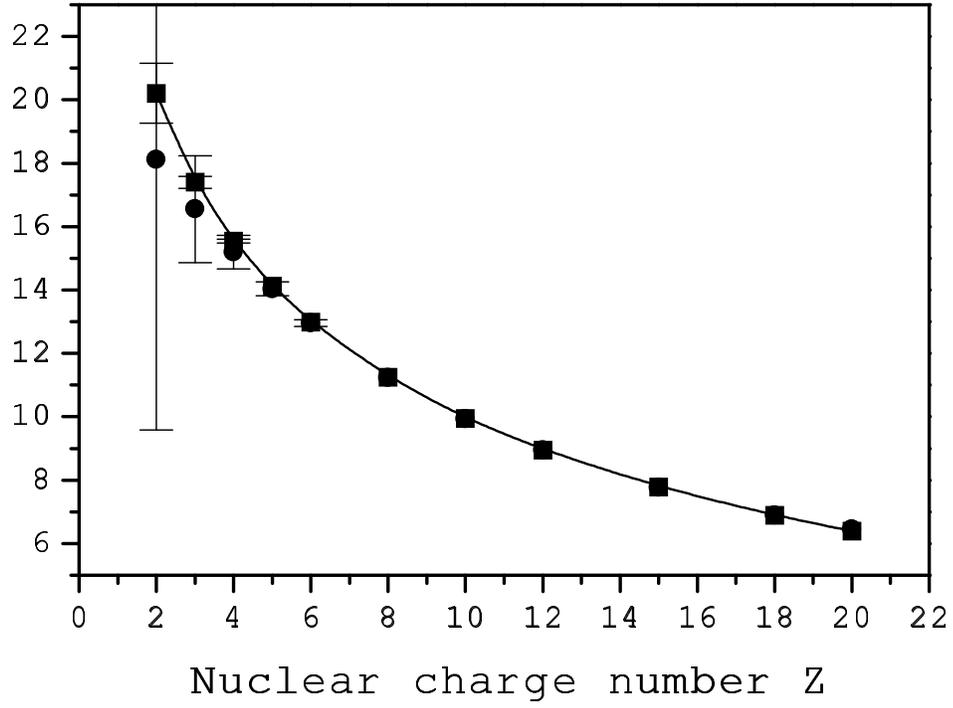}
}}
\caption{The higher-order contribution to the $1s$ electron $g$ factor for H-like ions
$F_{\rm h.o.}(Z)$,
defined by (\ref{eq6}). The squares indicate our numerical values and the circles stand
for the results of Beier {\it et al}.
}
\end{figure}

\end{document}